# Improving the clinical utility of lower-limb surface electromyography (sEMG) by quantifying and correcting for location changes in inter-session recordings

*Short title*: **HDsEMG robustness through location quantification**

Fraser Douglas[1,2], Mona Pei[1,2], Quoc Sy Vu[3], Linh Le[3] and Calvin Kuo[1,2]

[1] School of Biomedical Engineering, University of British Columbia, British Columbia, Canada
[2] Centre for Aging SMART, University of British Columbia, Canada
[3] Focal Lines Technologies, Vancouver, Canada

Corresponding Author:
Fraser Douglas
Centre for Aging SMART
2635 Laurel St.
Vancouver, BC
V5Z 1M9
Fraser.Douglas@ubc.ca

**Author Contributions**

Study conception: Fraser Douglas and Calvin Kuo; Data collection: Fraser Douglas and Mona Pei; Analysis: Fraser Douglas; Supervision: Calvin Kuo; Funding acquisition: Quoc Sy Vu, Linh Le, and Calvin Kuo; Writing: Fraser Douglas; Manuscript review: all authors.

**Abstract**

*Purpose*:

Surface electromyography (sEMG) can enable direct muscle activity measurement to support the recovery assessment of individuals with neurological and musculoskeletal disorders. Despite this, its broader adoption of sEMG has been limited given its sensitivity to changes in electrode location across sessions. To address this challenge and enable multi-session sEMG, this work develops a novel high-density sEMG (HDsEMG) algorithm to quantify changes in electrode location and mitigate its effects on common time and frequency domain sEMG features.

*Methods*:

11 healthy participants performed isometric and dynamic exercises with HDsEMG on four lower-limb muscles. These were repeated four times, reapplying arrays at shifted locations. The error between spatially-mapped HDsEMG metrics was then minimised to estimate the change in array location, with this compared against ground truth 3D scans. Lastly, relative feature differences across locations were computed at select electrodes to assess the degree to which inter-session sEMG effects were mitigated.

*Results*:

Electrode location estimates were improved over the assumption their location remained unchanged in 81.7% of cases, 37.6% identified within 1 cm of the ground truth. Feature differences computed between closest electrodes across locations per ground truth and algorithm estimates were statistically similar. Conversely, feature differences for the same electrode across locations were significantly greater, increasing the mean difference for the isometric max envelope amplitude from 15.9% with the algorithm to 21.1% without.

*Conclusions*:

The application of our algorithm effectively reduced inter-session feature differences arising from changes in electrode location. This approach can facilitate more direct cross-session feature comparisons, representing a promising step toward robust sEMG measurement for musculoskeletal and neurological recovery tracking.

## 1. Introduction

Many believe that surface electromyography (sEMG) possesses considerable promise in clinically assessing neurological and musculoskeletal health and recovery given the non-invasive window into muscle activity that it provides [1]. One such area that may benefit from the inclusion of sEMG is stroke rehabilitation. Specifically, features extracted from sEMG data have demonstrated potential in classifying impairment [2–4], as well as tracking recovery progress and informing rehabilitation recommendations [5–9].

Despite the promise shown by the technology, there exist a number of factors which have limited its clinical adoption; factors primarily revolving around the expertise required for its setup, use, and signal interpretation thereafter [10–12]. Central to this theme of expert knowledge for setup and use is the process of correctly locating the optimal recording site and placing the sEMG sensors to record from the muscle of interest. Given sEMG's signal properties are highly sensitive to changes in electrode location, consistent placement is crucial to ensure that changes in these properties arise primarily through physiological changes in the patient [13, 14].

The SENIAM guidelines are widely recommended to try and achieve consistent electrode placement and mitigate sensitivity to misplaced electrodes [15]. However, while these guidelines are comprehensive, their recommendations rely on the use of palpation and surface anatomical landmarks to identify the muscle belly for sensor placement. In healthy adults, this muscle belly identification process can be aided by muscle tone and the individual's ability to selectively tense the muscles of interest. However, in impaired populations who may lack this tone or voluntary contractile ability, the process can be far more complicated. In fact, related research has shown that in a transtibial amputee population, certified prosthetist orthotists were only able to identify the location of the muscle belly via palpation to within 1cm of the ultrasound-identified location in 28.8% of cases [16].

These are difficulties that are only exacerbated when the placement is performed by a non-expert or patient population. This is particularly relevant given that, as catalyzed by COVID-19, we have seen the traditionally in-person stroke recovery assessment sessions largely shift to instead being delivered remotely via videoconferencing platforms [17]. While beneficial to many given the increased accessibility afforded by these remote offerings, they rely on the subjective assessment of the patient's movement quality and muscle function through on-screen observation alone [18]. Thus, there exists an ongoing and concerted effort to enable more objective means of recovery quantification within these remote assessment sessions, recovery quantification that may be enabled through the information wearable sensors including sEMG can provide [19, 20].

In light of this, a critical question emerges: how can we mitigate the effects of electrode shift, both in and out of the clinical setting, to enable longitudinal sEMG assessment where changes in signal properties can be attributed to physiological changes in the patient? One area in which the effects of electrode shift have been explored considerably is that of myoelectric prosthesis control and gesture recognition. Here, researchers have proposed a variety of methods to achieve the same level of classifier performance post-shift as they saw pre-shift, be it through the signal features chosen for their classification model, or a means of adaptively recalibrating their classifier in the presence of shift [21–24]. Despite the merits of these approaches, they do not facilitate the comparison of feature values from the same muscle location across sessions, and are thus not applicable to the broad goal of applying sEMG in stroke recovery assessments.

There has, however, been some work instead aiming to resolve changes in electrode location by leveraging the increased spatial resolution afforded by high density sEMG (HDsEMG). Here, researchers applied independent component analysis (ICA), determined the major pattern, before using the shift of this pattern to understand whether the sensing array had moved [25]. While their results were promising, the physical electrode array was moved by a maximum distance of 1.4 cm, and only in the medial-distal, lateral-distal, medial-proximal, and lateral-proximal directions. Thinking of the sensor location variations that are expected clinically, consistently achieving a placement within 1.4 cm of the desired recording site is rare [16]. Additionally, upon recurring placements one would expect variations in the orientation of the sensor array alongside its translation, something that was not included in the ICA-based study.[16]

Thus, it is the pursuit of resolving these larger amplitude, both translational and rotational placement variations to mitigate errors in sEMG feature values from which the present work stems. In turn, two hypotheses emerge: 1) by leveraging the increased spatial resolution of HDsEMG, we hypothesise that we can develop an algorithm that accurately determines the change in sensor location between recording bouts, and 2) by using this algorithm to determine the closest electrodes across recording bouts we can mitigate the impact of electrode shift on sEMG features extracted from both isometric and dynamic exercise data.

## 2. Methods

The data used to explore these hypotheses came from 11 healthy participants (three males, eight females, age 28.75 ± 5.9). Participants provided informed consent in compliance with the Declaration of Helsinki under a protocol approved by the University of British Columbia Behavioural Research Ethics Board (H22-02423). The data collection protocol was first detailed in our article characterizing sEMG feature errors in lower limb muscles [14], with relevant details described below.

### *2.1. Instrumentation*

Data were collected from a variety of lower limb muscles of the dominant leg across participants. Specifically, these muscles were the gastrocnemius medialis (GM), tibialis anterior (TA), semitendinosus (ST), and tensor fascia latae (TFL). Muscles were chosen due to the differences they present when comparing stroke compensations to healthy gait [26–29]. Participants were allowed to opt out of ST and TFL instrumentation if they were uncomfortable with having their thigh instrumented.

Standard skin preparation was performed for the array and ground electrode sites: excess hair was shaved, the skin was abraded using Nuprep Skin Prep Gel, and the area was wiped clean prior to instrumentation. The 64-channel 8x8 1cm-spaced HDsEMG arrays (HD10MM0808, OT Bioelettronica, Fig. 1a) were fitted with their adhesive sheets before being prepared with conductive paste (Ten20 Conductive Paste, Weaver & Co.). Initial array placement was done with respect to SENIAM recommendations [15], while the ground electrodes were attached to electrically neutral tissue at the ankle for GM and TA recordings, and the knee for ST and TFL recordings. At most, two arrays were attached to the leg at any given time: one on the shank, and one on the thigh.

After array adhesion, a strip of hypoallergenic kinesiology tape was placed overtop the array to ensure it was secure. Next, markers were attached to the leg surrounding the array, as well as two on the array itself (Fig. 1a), as reference markers for localizing the array on the leg. Lastly, an elastic compression wrap was placed over the sensing instrument to maximise skin compliance during data collection. This compression wrap was also used to house the wireless amplifier (Sessantaquattro+, OT Bioelettronica) which was connected both to the array and ground electrode.

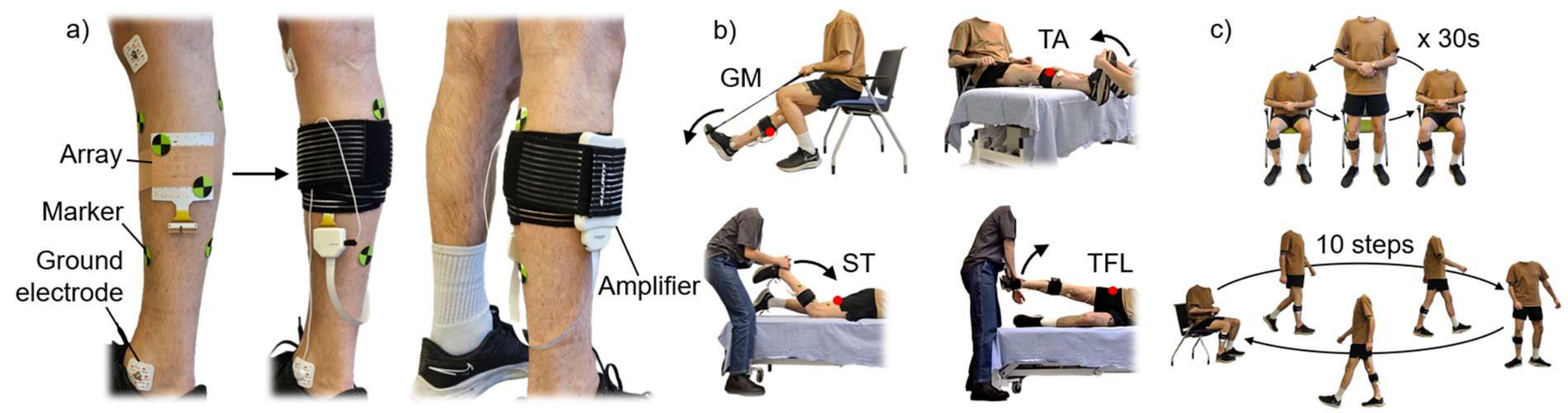

Figure 1. Pictorial representation of the instrumentation and HDsEMG data collection procedure with a) depicting the sensor instrumentation for the GM specifically, b) showing the isometric exercises performed for each muscle, and c) showing the dynamic exercises performed across muscles.

### *2.2. Data Collection*

Once the array was secured in its initial position, the participant performed 3 x 10s repetitions of each isometric exercise relevant to the instrumented muscles (Fig. 1b). For the GM this represented maximal plantarflexion against a strap while seated with the instrumented leg fully extended, for the TA the participant maximally dorsiflexed while the motion was resisted by the research assistant, for the ST the participant was prone with their knee bent to approximately 45 degrees maximally flexing their knee against the research assistant, and for the TFL they lay laterally and maximally abducted their hip while this motion was again resisted by the research assistant. These were followed by two dynamic exercises (Fig. 1c): 30s of sit-to-stand (STS) repetitions, and a 10-pace timed-up-and-go (TUG). These are two dynamic exercises often performed during stroke assessment sessions [30, 31]. HDsEMG data were recorded at a sample rate of 2000Hz. Following the set of exercises, the elastic wrap was removed and a 3D scan was taken of the leg with array still attached using a 3D scanner (Structure Sensor Pro, Structure) attached to an iPad (iPad Pro 5th-Generation, Apple). This was done to record the array's placement relative to the reference markers on the leg (Fig. 1a).

Post-scan, the array was translated by up to 4.25cm away from the muscle belly (location 0) to fall somewhere within location segment 1 (Fig. 2a). Additionally, the array was rotated about its center by between -30 and 30 degrees prior to reapplication. After the location shift had been performed and the array was reapplied, the elastic wrap was again placed overtop the array, and the exercises were reperformed. This process was repeated for all placement segments 1-3 (Fig. 2a). These location shifts acted as a representation of the imprecise placement that may be performed by non-expert populations when applying sEMG sensors, particularly relevant for its application out-of-clinic. Once EMG and scan data were collected for all of the segments (0-3), the array was removed from the leg and one final scan was performed with just the leg and the markers present (Fig. 2b).

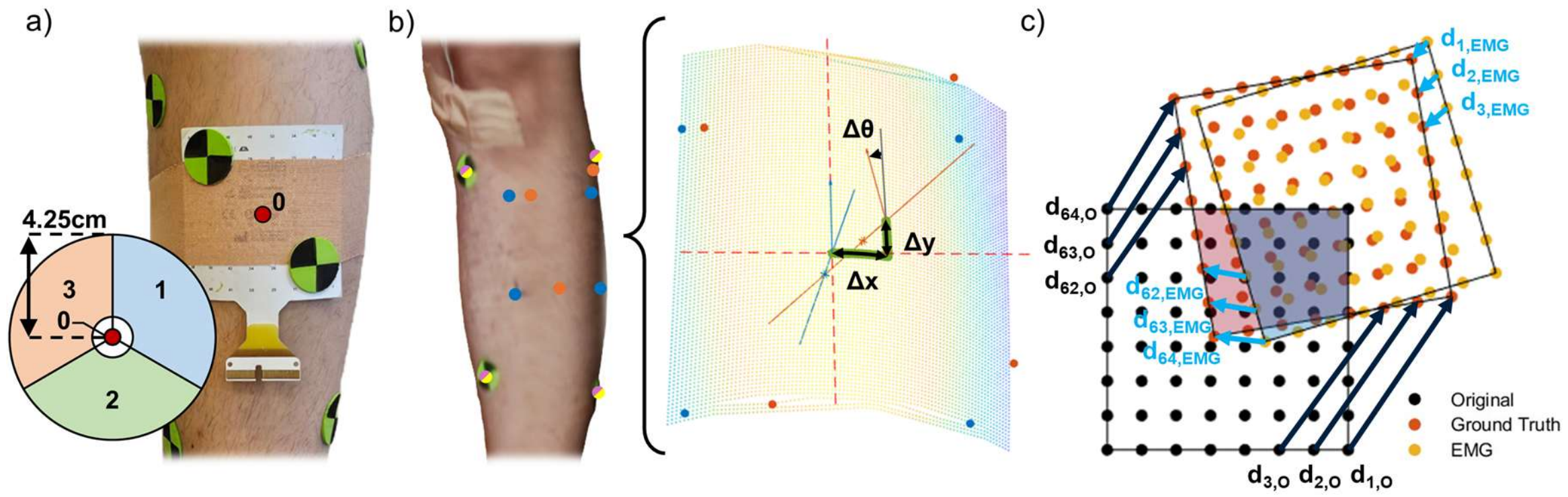

Figure 2. Visual showing a) the initial array placement as well as the placement segments representing how the array location was varied for exercise repetitions 0-3, b) a representation of how the registered scans were used to extract ground truth shift, and c) the process by which mean Euclidean distance was calculated between the case that the array was assumed to have not shifted and the EMG-estimated shift

### *2.3. Extraction of Ground Truth Shift from 3D Scans*

For location shift identification, three scans were used in each case: 1) the scan showing the array at its original location, 2) the scan of the array in its shifted location, and 3) the scan of the leg without the array present. By using the locations of the markers adhered to the skin, the array scans were registered to the no array scan. Next, using the locations of the arrays' corners, the array center locations were determined and projected onto the skin, before then identifying the transformation that must have occurred between the array's original and post-shift location in terms of "$x$", "$y$", and "$\theta$". Here, "$x$" is the distance over the skin parallel to the top and bottom edges of the array in its original location, "$y$" is the distance perpendicular to this, and "$\theta$" is the rotation about the center of the array following the translation (Fig. 2b). By knowing the location of the array as well as the configuration of its electrodes, this allowed for the extraction of each electrode's location. Using this information, we were are able to identify the Euclidean distance between each electrode before and after shift (Fig. 2c), averaging across the 64 electrodes to obtain the mean Euclidean distance (μED), a metric that will be used to assess the performance of the EMG location shift estimation algorithm.

### *2.4. EMG Location Change Estimation*

First, a pair of HDsEMG data signals from before and after a change in the sensor array location had occurred were fed through an $8^{th}$-order Butterworth bandpass filter with cutoff frequencies at 20Hz and 450Hz. Additionally, a series of infinite impulse response notch filters were implemented to remove mains interference at 60Hz and its harmonics. Separately, the signal envelopes were generated by rectifying the EMG signals and low-passing them at 2Hz using an $8^{th}$ order Butterworth filter. The post-notch filtered data were then segmented (Fig. 3b) into their individual contractions, with the isometric signals segmented by thresholding their signal envelopes at 20% of their maximum value. The TUG and STS data were initially segmented the same way as the isometric data, before having their contraction bounds adjusted manually by a single researcher through visual inspection of the data.

To estimate the change in sensor array location (Fig. 3), three metrics were extracted from all 64 channels of the post-notch filtered isometric data: 1) mean absolute pairwise correlation, 2) mean frequency, and 3) maximum envelope amplitude. The mean absolute pairwise correlation used Pearson's correlation coefficient as shown below (eqn. 1) where $r_a$ represents the mean absolute pairwise correlation of channel $a$, $C$ the total number of channels (64 in this case), $n$ the number of samples in the current contraction, $X_a$ the EMG signal of channel $a$ after filtering, where $\bar{X}_a$ is its mean value. $Y_b$ is the EMG data of the channel against which $X_a$'s correlation is being calculated, and $\bar{Y}_b$ is its mean value. This was chosen as it gives a spatial representation of regions of common underlying muscle activity within the HDsEMG array without being impacted by changes in signal amplitude across recordings.

$$r_a = \frac{1}{C}\sum_{b=1}^{C}\left|\frac{\sum_{i=1}^{n}\left(X_{a,i}-\bar{X}_a\right)\left(Y_{b,i}-\bar{Y}_b\right)}{\sqrt{\sum_{i=1}^{n}\left(X_{a,i}-\bar{X}_a\right)^2\sum_{j=1}^{n}\left(Y_{b,j}-\bar{Y}_b\right)^2}}\right| \tag{1}$$

To extract mean frequency from the filtered contractions, we first estimated their power spectral density (PSD), $S(f)$, using Welch's averaged periodogram with a 200ms Hamming window. The total power of each contraction, $P$, was estimated through the trapezoidal numerical integration of the PSD with frequency step, $\Delta f$, and bandwidth $f_{N+1} = Fs/2 = 1000Hz$ (eqn. 2).

$$P \approx \sum_{n=1}^{N}\Delta f\,\frac{\left(S(f_n)+S(f_{n+1})\right)}{2} \tag{2}$$

The mean frequency for each contraction, $MNF$, was then taken as the frequency in Hz where the average power across the spectrum was reached (eqn. 3). This was included as it provides information on the frequency content of the signal which is impacted by where on the muscle the electrode is positioned.

$$MNF \approx \frac{\sum_{n=1}^{N}\Delta f\,\frac{\left(f_nS(f_n)+f_{n+1}S(f_{n+1})\right)}{2}}{P} \tag{3}$$

For the max envelope amplitude, this simply represented the greatest value of the contraction's envelope, and represents a common time domain feature, thus presenting information not represented within the previous two metrics. For all three of these, they were then averaged across contractions within a trial.

Following feature extraction, a gradient descent-based global optimiser was configured using the Matlab 'GlobalSearch' function [32, 33]. The objective function (eqn. 4) was defined as the weighted sum of the root mean squared error (RMSE) between the three metric values (mean frequency $\{MNF\}$, max envelope value $\{env\}$, mean pairwise correlation $\{r\}$) of the $N_O$ electrodes in the overlapping region between the pre- and post-shift location as defined by the function's input. This input to the objective function was a 3-element vector containing the shift estimate in $x$, $y$, and $\theta$ (rotation), and thus the optimiser sought to identify the shift minimizing the weighted error between metrics. Looking at the objective function, $c_n(x, y, \theta)$ represents the location of the $n^{th}$ electrode in the overlapping region of the start position array as defined by $x, y, \theta$. The metric values at this location for the end position array were spline interpolated from the values at the surrounding electrodes for direct comparison.

$$\underset{x,y,\theta}{\operatorname{argmin}} \left( \begin{array}{c} w_1 * \sqrt{\frac{1}{N_O}\sum_{n=1}^{N_O}\left(MNF_2\left(c_n(x,y,\theta)\right) - MNF_1\left(c_n(x,y,\theta)\right)\right)^2} \\ +w_2 * \sqrt{\frac{1}{N_O}\sum_{c=1}^{N_O}\left(env_2\left(c_n(x,y,\theta)\right) - env_1\left(c_n(x,y,\theta)\right)\right)^2} \\ +w_3 * \sqrt{\frac{1}{N_O}\sum_{c=1}^{N_O}\left(r_2\left(c_n(x,y,\theta)\right) - r_1\left(c_n(x,y,\theta)\right)\right)^2} \end{array} \right) \tag{4}$$

It should be noted that, for the correlation, the overlapping region values were calculated using only the information from electrodes in the overlapping region (Fig. 3c,d). Additionally, for the mean frequency and max envelope values, prior to being fed into the optimiser they were min-max normalized such that the smallest value across the shift pair was zero, and the highest was one. In doing so, all three metrics had values of similar amplitude, ranging from zero to one.

For weight determination, a subset of 30 shift pairs were randomly selected from the dataset, and the optimiser was run on their data with equal weights ($w_1$, $w_2$, $w_3$) for the correlation, mean frequency, and max envelope amplitude. Using the mean RMSE of each feature, the weights were then calculated as the value that multiplied the mean RMSE to equal that of whichever feature had the highest mean RMSE. This resulted in a weight of 2.72 ($w_1$), 1.24 ($w_2$), and 1.00 ($w_3$) for the correlation, mean frequency, and max envelope amplitude RMSEs respectively.

With these weights, the shift estimation algorithm (Fig. 3) was then run on the remaining shift pairs. To assess the performance of the algorithm, the mean Euclidean distance was calculated between where the electrodes were thought to be per the shift algorithm, and where they actually were per the ground truth electrode location as extracted from the 3D scans (Fig. 2c). These were then compared to mean Euclidean distance from the origin, representing the assumption used currently that no shift occurred between recording bouts.

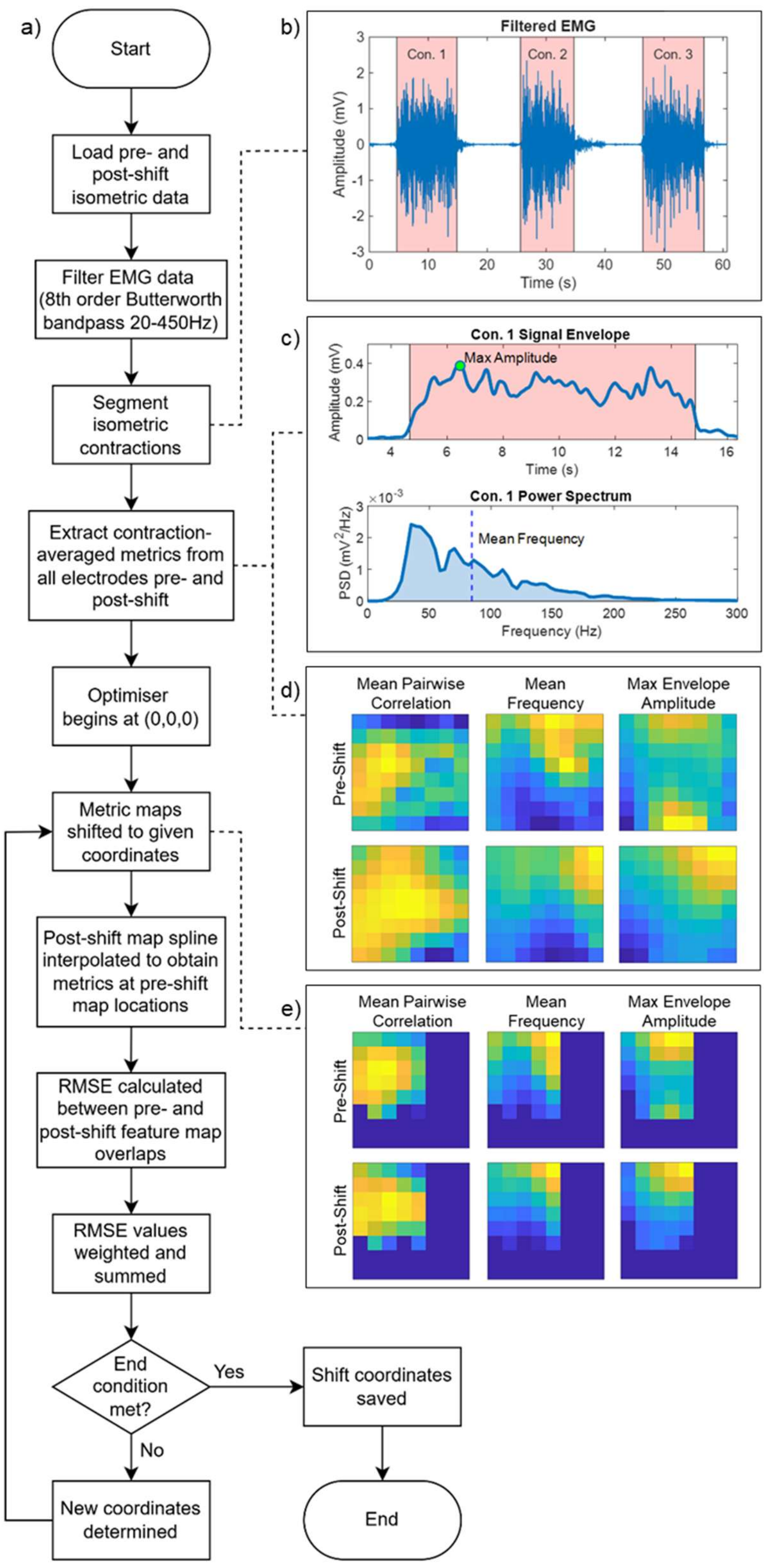

Figure 3. a) Flow diagram representing the process followed by the HDsEMG location shift estimation algorithm with b) signal filtering and segmentation, c) signal metric extraction and d) map formation, and e) location shift identification visualized

*2.4. EMG Feature Analysis*

With the overarching goal of this work being to reduce the effect of shift on sEMG features thus supporting longitudinal EMG-based recovery tracking, this final step seeks to understand how the shift algorithm can reduce feature variability across recording bouts. Furthermore, given that the recovery assessment sessions generally involve the patient performing dynamic exercises, the following analysis was performed for both the isometric recordings as well as the dynamic exercise recordings.

For this analysis, we compared five features before and after shift in three conditions (Fig. 4): 1) the closest electrodes before and after a change in recording location per the ground truth 3D scans, 2) the closest electrodes before and after a change in recording location per the HDsEMG algorithm, and 3) the same electrodes before and after the change in recording location using the four electrodes identified by condition 1 and 2 – this is analogous to what is currently done, comparing signal information directly across recordings without accounting for a change in location. For conditions 1 and 2, the features were first compared across the closest locations before and after shift (i.e. across two repetitions) by computing their relative percent difference.

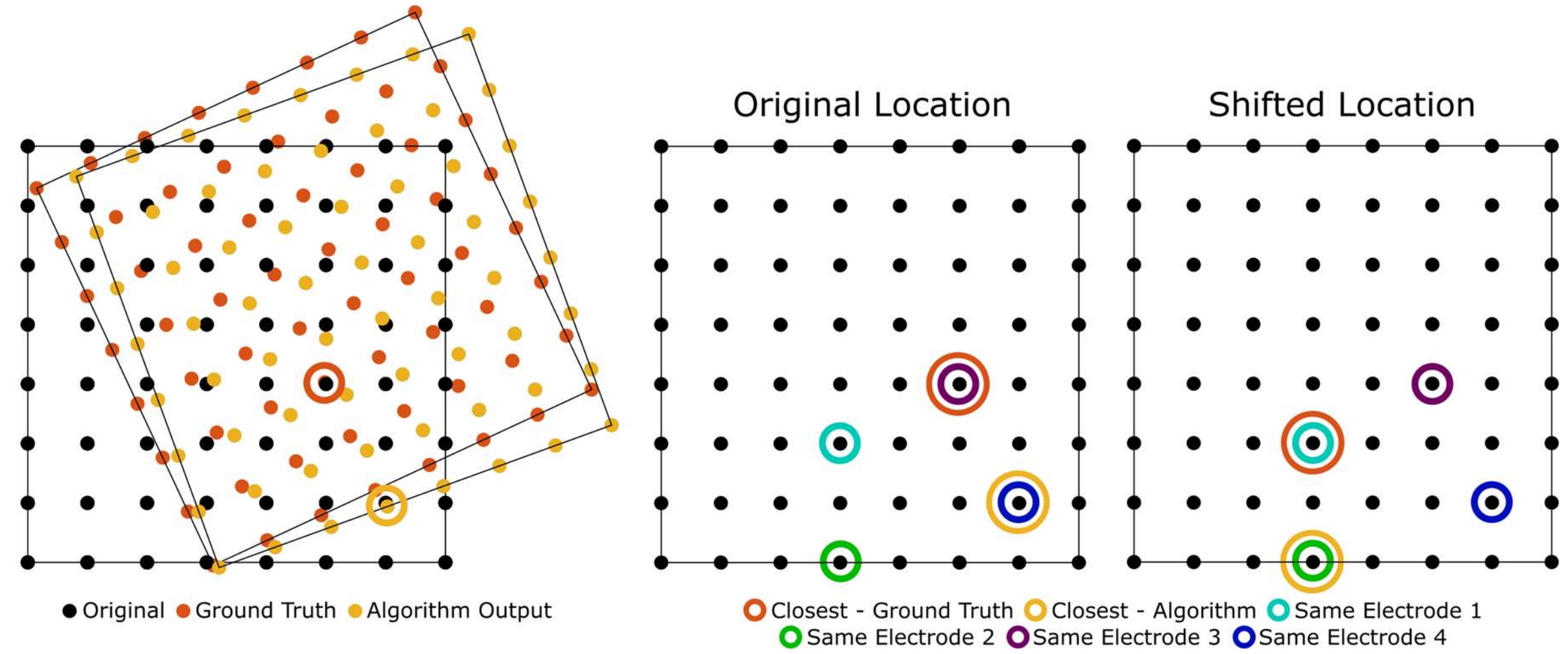

Figure 4. Visual representation of the electrode pair comparisons within the feature analysis. The left panel shows the array's original location, as well as its updated location as identified by the 3D scans and HDsEMG algorithm; circled are the closest electrodes for each of these conditions. The right panel shows the original and shifted location arrays, with coloured circles denoting the electrodes used for the feature comparisons – these are the closest electrodes per ground truth, closest per the algorithm, and then the same electrodes involved in these first conditions before and after the change in location.

The relative difference ($d_r$) was taken as the absolute value of the difference between the feature value at the electrode in the shifted location ($F_B$) and the electrode in the original location ($F_A$) normalized by the feature value for the electrode in the original location ($F_A$) (eqn. 4). This was done for each electrode pair (Fig. 4).

$$d_r = 100\% * \left|\frac{F_B - F_A}{F_A}\right| \tag{4}$$

Then, by using the closest electrodes across all locations (i.e. across all four repetitions) we calculated the features' mean-normalized standard deviation, more closely representing the projected real world use wherein features are compared longitudinally across numerous sessions.

The mean-normalized standard deviation ($S_\mu$) was taken as the standard deviation of the contraction-averaged-feature values at the electrodes of interest ($F$) across $N_s$ recording bouts normalized by the mean contraction-averaged feature value across repetitions at these same electrodes ($\bar{F}$) expressed as a percentage (eqn. 5).

$$S_\mu = \frac{100\%}{\bar{F}} * \sqrt{\frac{1}{N_s - 1}\sum_{i=1}^{N_s}|F_i - \bar{F}|^2} \tag{5}$$

For these comparisons, we extracted the mean (eqn. 3), median, and peak frequencies from the bandpassed and notch-filtered contractions, features used previously to quantify differences between healthy and stroke populations [34]. Specifically, a reduction in the mean and median frequency has been associated with increased muscle fatigue [7, 35, 36], a common affliction in stroke patients [37–39]. Additionally, we extracted the integrated EMG (iEMG) value from the contraction envelopes. Increases in iEMG suggest improvements in firing rate and motor unit recruitment, both of which can be linked to stroke recovery [40, 41]. Finally, we examined the signal envelope's maximum amplitude as described in section 2.4. which correlates to muscle force production and thus may indicate changes in stroke patients' physical ability [42, 43].

The median frequency for each contraction, $MDF$, measured in Hz, was then defined as the frequency at which 50% of the total power (eqn. 2) within the contraction's power spectrum was reached (eqn. 6).

$$0.5\,P \approx \sum_{n=0}^{M} \Delta f \frac{\left(S(f_n) + S(f_{n+1})\right)}{2} \tag{6}$$
$$MDF = f_{M+1}$$

For peak frequency, this was simply taken as the frequency value which showed the highest power within each contraction. Finally, for the iEMG, this was taken as the trapezoidal numeric integral of the signal envelope for each contraction. Then, for each of the obtained features, we averaged their values across all contractions within each exercise repetition. Note that we did not perform signal normalization for our time-domain features.

## 3. Results

### *3.1. EMG Location Shift Quantification Algorithm*

Of the 293 location pairs for this analysis, 30 were used for weight calculation. When run across the 263 remaining location pairs for the isometric exercise data, the algorithm improved the µED over assuming the position remained unchanged (origin assumption) in 81.7% of cases (Fig. 5, left). The median µEDs under the origin assumption were 3.16cm, 2.93cm, 2.79cm, and 2.82cm, being improved to 1.27cm, 0.99cm, 1.42cm, and 1.22cm by the shift algorithm for the GM, TA,

ST, and TFL respectively. Across all shift pairs, the median μED was improved from 3.00cm to 1.22cm (95%CI: 1.08cm to 1.36cm). For the unimproved results, 42 shift estimates were worsened while no difference was seen between the origin assumption and shift algorithm for 6 shift pairs. Of the 40 that were worsened, the average μED increase was 1.27cm.

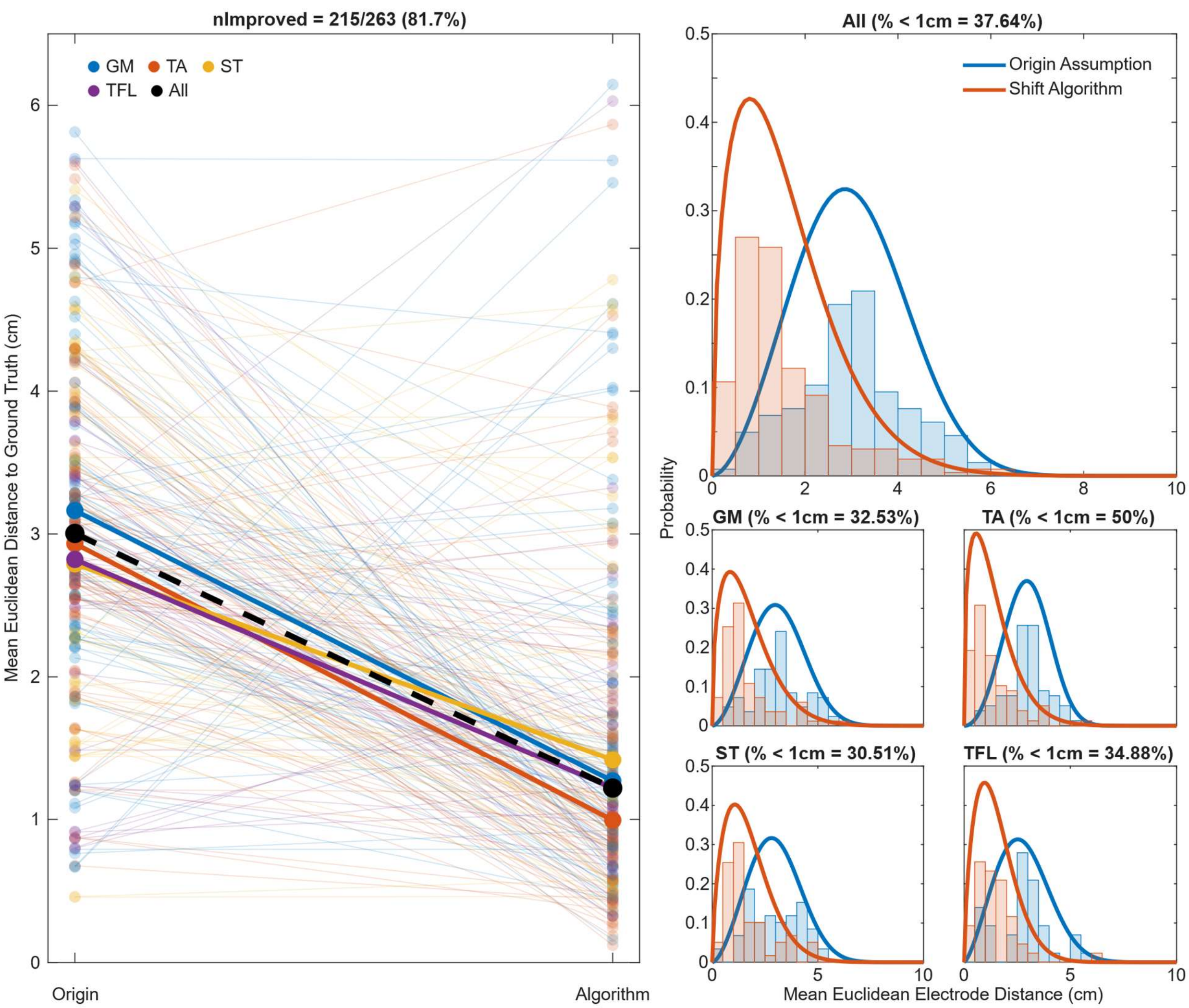

Figure 5. Isometric results from the application of the EMG location shift algorithm. Left: mean euclidean distance to the ground truth position in the scenario where no shift is assumed (origin, left) and when the shift algorithm is applied (algorithm, right). Narrow lines show the individual shift pair results for each muscle, thick lines show the median, the black dashed line shows the median across all muscles' results. Right: histograms showing the distribution of mean Euclidean distances to the ground truth for the origin assumption and shift algorithm outputs. Solid lines represent Weibull distributions fit to the data.

Additionally, when looking at the μED distribution results (Fig. 5, right) we get a better sense of the improvement where 37.6% of all shifts were identified to within 1cm; this is equal to the HDsEMG array's inter-electrode distance. This also presents the differences across muscles, where 50.0% of shifts were detected to within 1cm for the TA as compared to 32.5% 30.5% and 34.9% for the GM, ST, and TFL respectively.

*3.2. Feature Analysis*

From the feature analysis of the isometric exercises (Fig. 6) we see that for all features, with the exception of peak frequency, both the relative difference across shift pairs and multi-session mean-normalized standard deviation is decreased in the closest GT and EMG algorithm cases as compared to the same electrode. Furthermore, per the Wilcoxon rank sum test results, in all cases outside of the peak frequency, the same electrode results were significantly different to both the closest GT and EMG algorithm results. Additionally, across features the closest GT and EMG algorithm results passed the Wilcoxon rank sum test under the hypothesis that their data are samples from continuous distributions with equal medians.

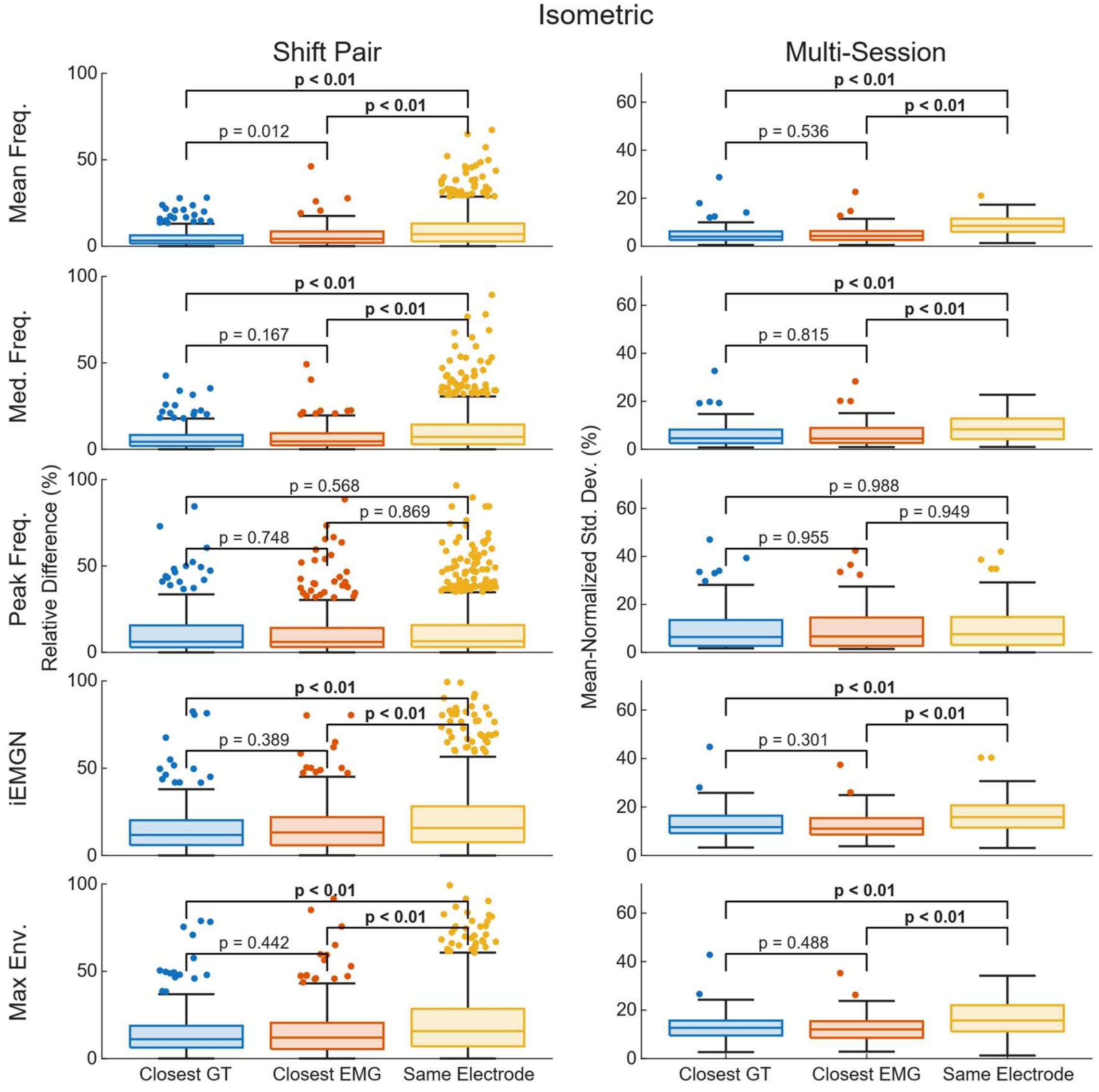

Figure 6. Feature analysis results across individual shift pairs (left) and all recording bouts (right) for the isometric exercises across all muscles. Individual axes present the results for the closest electrodes per the ground truth 3D scans (Closest GT), closest electrodes per the shift algorithm (Closest EMG), and the same electrodes before and after shift. Brackets present the results from the Wilcoxon rank sum tests with the null hypothesis that the data are samples from continuous distributions with equal medians.

For the dynamic features (Fig. 7), although the difference between the closest GT and EMG algorithm distributions as compared to the same electrodes differ less in magnitude than for the isometric results, we see similar trends with the statistical significance for the shift pair results here. In this regard, we again see that the closest GT and EMG algorithm results are significantly different than those of the same electrodes for most features (this time with the exception of both the peak frequency and integrated EMG) while being statistically similar to each other. In contrast, we see no results rejecting the null hypothesis in the case of the multi-session analysis, and in fact for the peak frequency, integrated EMG, and max envelope values we see a lower normalized standard deviation for the same electrode case as compared to the closest EMG algorithm results.

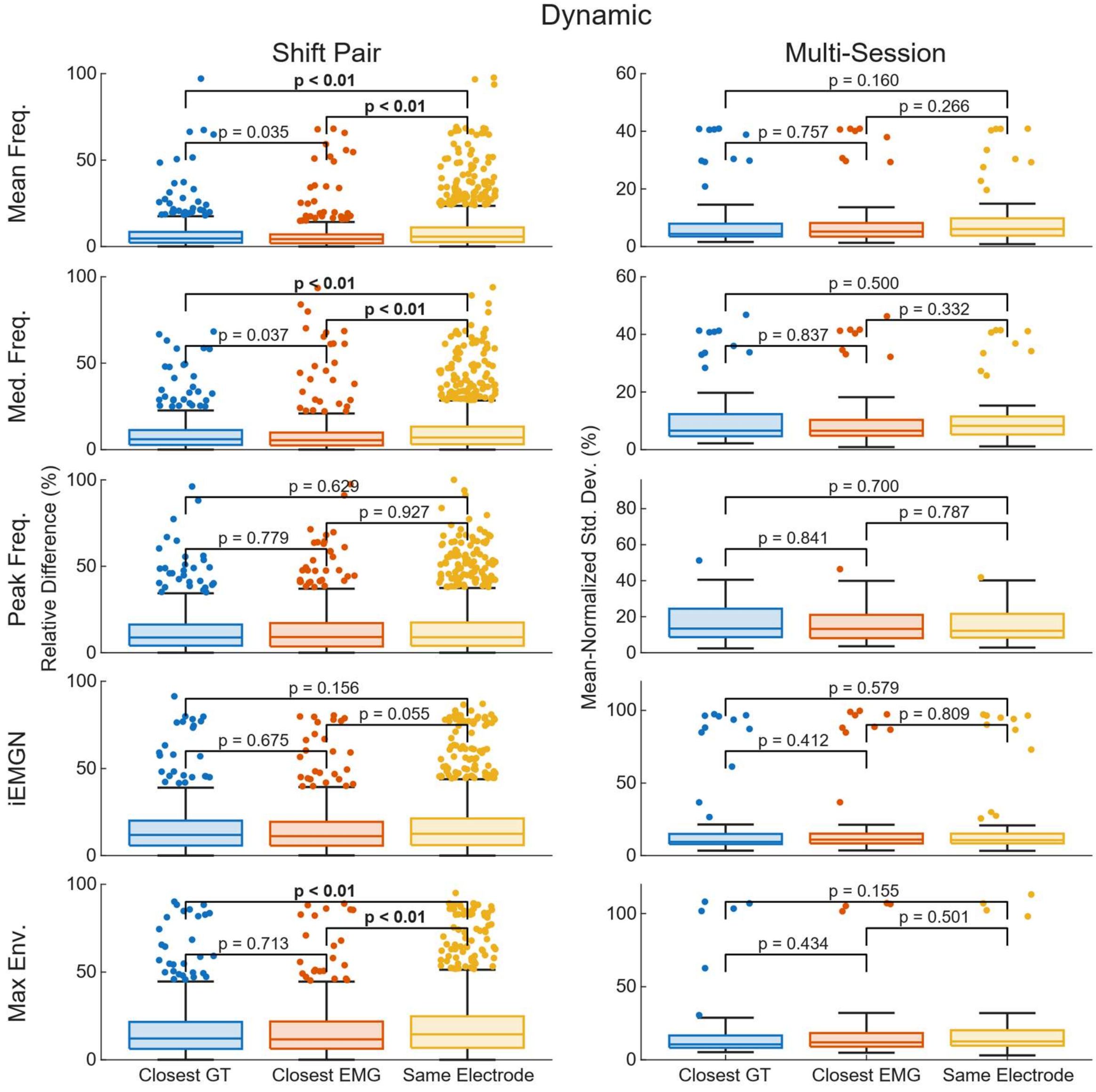

Figure 7. Feature analysis results across individual shift pairs (left) and all recording bouts (right) for the dynamic exercises across all muscles. Individual axes present the results for the closest electrodes per the ground truth 3D scans (Closest GT), closest electrodes per the shift algorithm (Closest EMG), and the same electrodes before and after shift. Brackets present the results from the Wilcoxon rank sum tests with the null hypothesis that the data are samples from continuous distributions with equal medians.

## 4. Discussion

Through this work, we have shown that we can effectively estimate the change in location of HDsEMG electrodes and use this to reduce the inter-session variability of extracted features. This will enable the robust use of sEMG in longitudinal assessment settings as we have demonstrated how using the algorithm to estimate the two closest electrodes before and after a change in location produces feature differences which are statistically similar to comparing the two closest electrodes per the ground truth 3D scans.

Regarding the algorithm's location estimation, typically we were able to identify the change in physical location of the HDsEMG array to within 1.22cm of its true value using EMG data alone, and in over 80% of cases the algorithm improved the estimate of sensor location over the assumption used presently which is that its location remained unchanged. Furthermore, through the use of the algorithm, the mean Euclidean distance between the electrodes' true locations and those of the shift estimate was identified to within 1cm in 37.6% of cases as compared to less than 5% for the origin assumption. This is significant given that previous work asking certified prosthetist orthotists to identify the muscle belly location on transtibial amputees only saw the correct location being identified to within 1cm in 28.8% of cases [16]. Thus, by extrapolating from the present work, this implies that the algorithm may be used to consistently record data from the same location on a muscle across sessions at a precision perhaps greater than what would be achieved through clinician-guided placement.

Beyond extracting the shift itself, we were able to demonstrate that by using the algorithm's outputs to estimate the closest electrodes across recording bouts, the feature differences were statistically similar to those from determining the closest electrodes using the 3D scans. This trend was consistent for both the isometric and dynamic exercises, demonstrating how the use of the algorithm could improve the reliability of multi-session feature comparisons. When comparing these to the results obtained when using the same electrodes before and after shift, these were significantly larger than both sets of closest electrode results in the isometric case (other than for peak frequency). These statistical differences remained for the shift pairs in the dynamic case (other than iEMG), although no significant differences were seen in the dynamic multi-session results wherein features are compared across all shifts.

When viewing these feature results, it is important to remember the factors at play that can lead to the differences observed: changes in location, changes in the skin-electrode interface, and changes in exercise effort or fatigue. While the closest electrode results are minimally impacted by changes in location ($0.12\pm0.10$cm between closest electrodes per ground truth), the physical electrodes being compared across recording bouts will have different conductive properties (e.g. more or less conductive gel, or a difference in electrode conductance itself) which can contribute to the observed feature differences. Although the impact of these changes has been shown to be less than that of electrode shift [14], this would not impact the same electrode results in the present study given that their conductive properties should remain consistent across recordings. However, in the projected use case wherein comparisons are made across entirely separate recording sessions, comparing the same electrode across sessions would compound the errors associated with both a change in location and conductive properties. Furthermore, our previous work has shown that this effect is more prevalent for the time domain features, which can go some way towards explaining the closer, at times lower, error distributions for the dynamic exercises within these features specifically.

In addition, we hypothesise that, given the movement of the muscle underneath the sensing array during the dynamic exercises [44], this can lead to an almost averaging effect wherein each electrode sees activity from a variety of muscle locations during the contractions. This can further reinforce why we see less statistical significance when analysing these exercises specifically. To address this and further support the analysis of dynamic exercise data, we hope to develop a temporal shift algorithm building on the present work to, rather than solely determining shift across recording bouts, track the motion of the muscle itself during dynamic exercises. This would facilitate the extraction and comparison of features at the same point within the muscle at all times.

On the note of dynamic exercises, it is also important to acknowledge that, despite the shift algorithm performing well on the isometric data, when applied to the dynamic data the success is varied. Here, the algorithm improved the shift estimate in 63.1% and 56.3% of cases, identifying the shift to within 1cm of the ground truth in 17.3% and 15.5% of cases for the STS and TUG respectively. What was promising however, was that for the STS data, the shift for the TA specifically was detected to within 1cm in 32.3% of cases. We believe the poorer shift performance here can be attributed to the previously-discussed movement of the muscles, at times imprecise manual contraction segmentation, and crosstalk from surrounding musculature. While this movement is less of an issue for the isometric exercises, crosstalk can additionally explain the discrepancy in success between the TA and other muscles. Due to the fairly isolated nature of the TA and presence of the tibia adjacent to the muscle essentially acting as an electrical deadzone, these can provide valuable spatial landmarks for the determination of array shift.

Still, thinking to the eventual goal of using this technology in stroke recovery assessments, the success achieved for the TA specifically bodes well. This is due to the fact that, post stroke, drop foot is a common affliction as characterised by the inability to sufficiently dorsiflex, leading to difficulties clearing the foot during the swing phase of walking [45]. Thus, given the TA's role in this dorsiflexion, tracking the improvement of the muscle's contractile ability can be especially valuable in understanding the patient's functional recovery. Furthermore, the success of shift identification in the TA specifically during dynamic exercise provides hope that, with refinement, the shift algorithm might achieve similar performance across exercises – both dynamic and isometric. This is significant as isometric exercises are uncommon in standard stroke recovery assessments, and so to maximise the technology's clinical readiness, all methods should be applicable to existing workflows.

Despite the work's success, it is important for us to highlight some of the limitations. One key limitation is the nature of the shift; in this study all shifts were performed within a single, 3-hour data collection session rather than across days or weeks. As such, we are yet to confirm whether the algorithm achieves similar performance across recordings separated by, at times in the case of stroke assessment sessions, multiple weeks. That said, given that the algorithm is most dependent on signal correlation which represents regions of common electrical activity originating from the muscle's innervation zones before propagating through the surrounding fibers, these should remain consistent to support true multi-session shift quantification. Additionally, we must acknowledge that the healthy population recruited here is not representative of the stroke population we hope this technology can eventually support. To address these limitations, we plan to perform data collection recruiting an older chronic stroke population using similar methods to those presented here with the addition of a second recording session two weeks following the initial session. Using this data, we hope to explore the generalizability of the present findings both to a stroke population within a single session, as well as across multiple weeks.

With the algorithm itself, its limitations should also be highlighted. Firstly, given that it is a global optimisation-based approach, we cannot say with full confidence whether the shifts it identified truly represent the global minima in all cases. The reasoning for this can be seen when observing the optimisation surface (Fig. 8), where despite its broad shape showing a general region of minimum RMSE, it exhibits a significant amount of noise which leads to a myriad local minima. Thus, it is possible that should the optimiser's parameters be tuned further to more reliably determine the global minimum, algorithm performance may similarly improve beyond the results presented here.

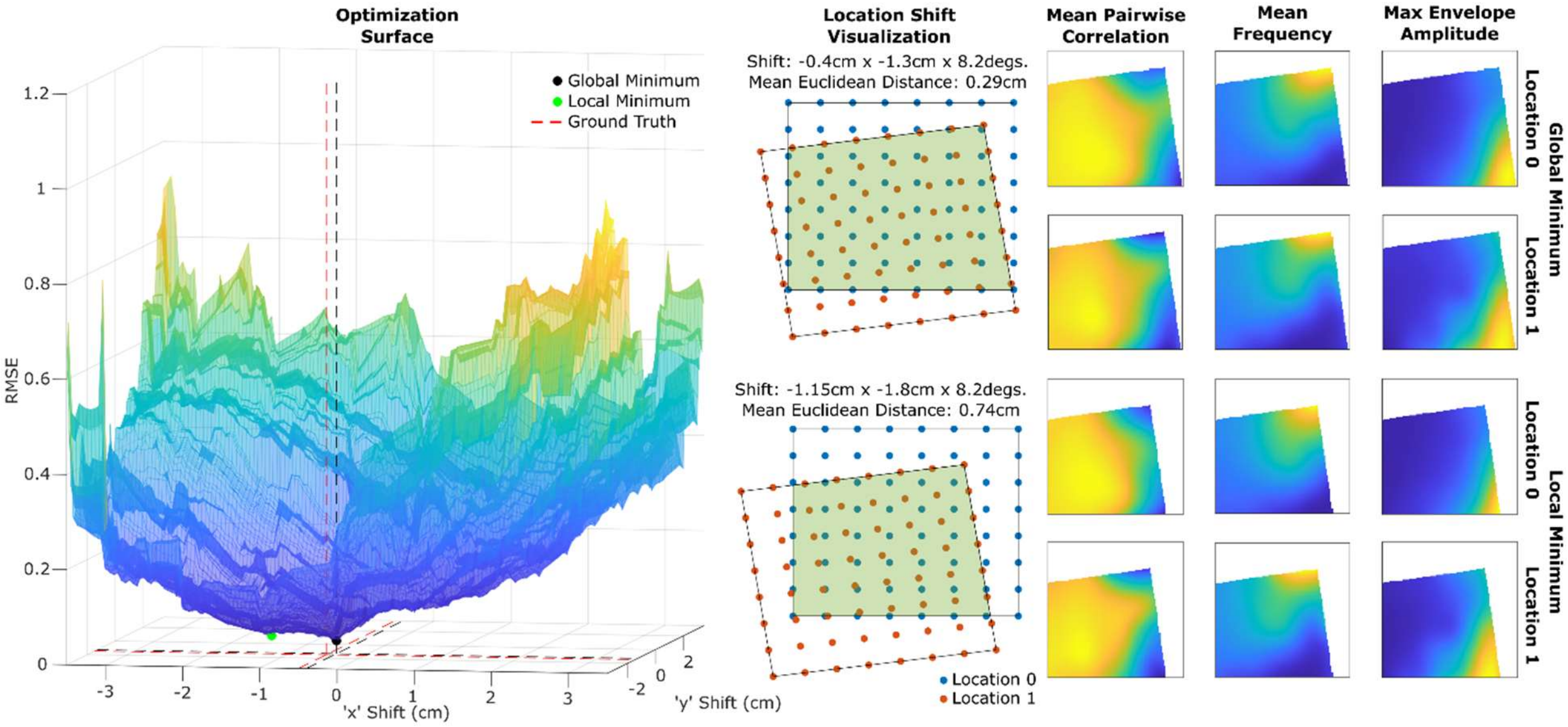

Figure 8. Sample optimization surface (left) showing the objective function's RMSE values for a single rotational shift at 0.05cm resolution in the 'x' and 'y' direction with the global minimum, local minimum, and ground truth shown. Location change visualization (middle) and metric maps from both locations for the global and local minima (right).

Next, one of our key objectives was to determine location changes more akin to those that one would expect through imprecise sensor placement. While we believe we were largely successful in doing so, the optimiser's search region was still bounded. Here, there was a maximum allowable shift of ±4cm in x, and y directions, and ±30° rotation. Referencing the transtibial EMG placement study, of the 36 clinician placements performed, 35 existed within the search region explored here, and thus we feel our bounds are justified [16]. These bounds were put in place because the location algorithm relies on there being some degree of physical overlap in array position across recording bouts, and so at its limits this represents a 17% overlapping area. From the shift results (Fig. 5), we see that even when the shift is close to these limits, the algorithm can still perform well, and so it is conceivable that an even lower overlapping area may still result in a resolvable shift. Similarly, should the inter-electrode spacing of the HDsEMG array be increased, it is possible that this may in turn increase the limits at which shift can be resolved, although it is reasonable to expect that this may be at the cost of resolution and so the percentage of shifts detected to within 1cm is likely to decrease.

Lastly on the algorithm itself, it determines the change in location using three normalized metrics (correlation, mean frequency, and max envelope amplitude), before we then used the shift results to explore the differences in feature values at the closest electrodes. Thus, we were concerned this was an almost cyclical process where the feature results may be biased by the working principle of the algorithm. To explore this further, we compared the closest electrodes as determined by the 3D scans to the electrodes with the lowest difference in feature values before and after the change in location. For the mean frequency, only 6.3% of pre and post shift electrodes were within 2cm of the closest electrodes as found using the 3D scans, with this decreasing to 3.4% for the max envelope amplitude. In light of this, we can say confidently that simply looking for the closest features across recording sessions is not a surrogate for resolving the change in location. Furthermore, as determined by the analysis of our testing set of 30 location pairs, the objective function is most dependent on the RMSE of the correlation maps (weight of 2.72 versus 1.24 and 1). When using this metric alone 70.3% of shift estimates were improved with 25.2% detected to within 1cm, and so the additional metrics simply act to enhance performance.

In conclusion, this work represents a significant step towards robust, inter-session sEMG analysis. This is enabled by a novel, gradient-descent based global optimisation algorithm that improves our understanding of electrode location, resolving physical electrode shift to less than 1cm 36.3% of the time. After correcting for the electrode location, sEMG features differences from electrodes at the same estimated location are statistically similar to feature differences at the closest electrodes identified through our ground truth 3D placement scans for isometric exercises, as well as dynamic exercises common to stroke recovery assessment sessions. Moving forward, this can potentially enable us to robustly track rehabilitation progress in stroke patients by quantifying muscle function with sEMG measures that are robust to changes in sensor location across recording sessions.

**Acknowledgements**

The authors would like to acknowledge that this research was conducted at the University of British Columbia Point Grey campus, which sits on the traditional, ancestral, and unceded territory of the xwməθkwəy̓əm (Musqueam) People. The authors would also like to acknowledge Ryan Lo for his assistance in data segmentation.

**Data Availability**

Participant data and Matlab code used in this paper are available at the following Dataverse link: https://borealisdata.ca/previewurl.xhtml?token=cea13433-d5a2-4e86-a30a-63b67083037e

## Declarations

Funding Declaration

This work was supported in part by a Canada Foundation for Innovation, John R. Evans Leaders Fund Grant, and a Canadian Natural Sciences and Engineering Research Council Discovery Grant awarded to Calvin Kuo, an AGE-WELL Graduate Award to Fraser Douglas, and a Michael Smith Health Research BC-Mitacs Accelerate grant with industry partners Focal Lines Technologies.

Conflict of Interest

The authors declare that this research was partially funded by Focal Lines Technologies, through partnership with MITACS. Authors LL and QSV are affiliated with Focal Lines Technologies. All other authors declare no competing financial interest. The funder had no influence on data analysis, interpretation or the decision to submit the manuscripts for publication.

Ethics Approval

Participants provided informed consent in compliance with the Declaration of Helsinki under a protocol approved by the University of British Columbia Behavioural Research Ethics Board (H22-02423).

## References

1. Al-Ayyad, M., H. A. Owida, R. De Fazio, B. Al-Naami, and P. Visconti. Electromyography Monitoring Systems in Rehabilitation: A Review of Clinical Applications, Wearable Devices and Signal Acquisition Methodologies. *Electronics 2023, Vol. 12, Page 1520* 12:1520, 2023. https://doi.org/10.3390/ELECTRONICS12071520

2. Al-Ayyad, M., H. A. Owida, R. De Fazio, B. Al-Naami, and P. Visconti. Electromyography Monitoring Systems in Rehabilitation: A Review of Clinical Applications, Wearable Devices and Signal Acquisition Methodologies. *Electronics 2023, Vol. 12, Page 1520* 12:1520, 2023. https://doi.org/10.3390/ELECTRONICS12071520

3. Roy, S. H., M. S. Cheng, S. S. Chang, J. Moore, G. De Luca, S. H. Nawab, and C. J. De Luca. A combined sEMG and accelerometer system for monitoring functional activity in stroke. *IEEE Transactions on Neural Systems and Rehabilitation Engineering* 17:585–594, 2009. https://doi.org/10.1109/TNSRE.2009.2036615

4. Kisiel-Sajewicz, K., Y. Fang, K. Hrovat, G. H. Yue, V. Siemionow, C. K. Sun, A. Jaskólska, A. Jaskólski, V. Sahgal, and J. J. Daly. Weakening of synergist muscle coupling during reaching movement in stroke patients. *Neurorehabil. Neural Repair* 25:359–368, 2011. https://doi.org/10.1177/1545968310388665/ASSET/IMAGES/LARGE/10.1177_1545968310388665-FIG5.JPEG

5. Steele, K. M., C. Papazian, and H. A. Feldner. Muscle Activity After Stroke: Perspectives on Deploying Surface Electromyography in Acute Care. *Front. Neurol.* 11:576757, 2020. https://doi.org/10.3389/FNEUR.2020.576757/BIBTEX

6. Klein, C. S., S. Li, X. Hu, and X. Li. Editorial: Electromyography (EMG) Techniques for the Assessment and Rehabilitation of Motor Impairment Following Stroke. *Front. Neurol.* 9:1122, 2018. https://doi.org/10.3389/FNEUR.2018.01122

7. Phinyomark, A., S. Thongpanja, H. Hu, P. Phukpattaranont, C. Limsakul, A. Phinyomark, S. Thongpanja, H. Hu, P. Phukpattaranont, and C. Limsakul. The Usefulness of Mean and Median Frequencies in Electromyography Analysis. *Computational Intelligence in Electromyography Analysis - A Perspective on Current Applications and Future Challenges* , 2012. https://doi.org/10.5772/50639

8. Cumming, T. B., M. Packer, S. F. Kramer, and C. English. The prevalence of fatigue after stroke: A systematic review and meta-analysis. *Int. J. Stroke* 11:968–977, 2016. https://doi.org/10.1177/1747493016669861

9. De Groot, M. H., S. J. Phillips, and G. A. Eskes. Fatigue associated with stroke and other neurologic conditions: implications for stroke rehabilitation. *Arch. Phys. Med. Rehabil.* 84:1714–1720, 2003. https://doi.org/10.1053/S0003-9993(03)00346-0

10. Campanini, I., C. Disselhorst-Klug, W. Z. Rymer, and R. Merletti. Surface EMG in Clinical Assessment and Neurorehabilitation: Barriers Limiting Its Use. *Front. Neurol.* 0:934, 2020. https://doi.org/10.3389/FNEUR.2020.00934

11. Manca, A., A. Cereatti, L. Bar-On, A. Botter, U. Della Croce, M. Knaflitz, N. A. Maffiuletti, D. Mazzoli, A. Merlo, S. Roatta, A. Turolla, and F. Deriu. A Survey on the Use and Barriers of Surface Electromyography in Neurorehabilitation. *Front. Neurol.* 11:1137, 2020. https://doi.org/10.3389/fneur.2020.573616

12. Feldner, H. A., D. Howell, V. E. Kelly, S. W. McCoy, and K. M. Steele. "Look, Your Muscles Are Firing!": A Qualitative Study of Clinician Perspectives on the Use of Surface Electromyography in Neurorehabilitation. *Arch. Phys. Med. Rehabil.* 100:663–675, 2019. https://doi.org/10.1016/J.APMR.2018.09.120

13. Wong, Y. M., and G. Y. F. Ng. Surface electrode placement affects the EMG recordings of the quadriceps muscles. *Physical Therapy in Sport* 7:122–127, 2006. https://doi.org/10.1016/J.PTSP.2006.03.006

14. Douglas, F., M. Pei, and C. Kuo. Characterizing sEMG Feature Errors in Lower Limb Muscles Due to Electrode Location and Skin–Electrode Interface Changes. *IEEE Trans. Instrum. Meas.* 74:, 2025. https://doi.org/10.1109/TIM.2025.3568954

15. SENIAM. , 2022.at <http://www.seniam.org/>

16. Rostamjoud, F., F. B. Porkelsdottir, A. O. Sverrisson, S. Brynjolfsson, and K. Briem. Improving Electromyography Electrode Placement Accuracy in Transtibial Amputees: A Comparative Study of Ultrasound and Palpation Methods. *IEEE Trans. Neural Syst. Rehabil. Eng.* PP:133–139, 2024. https://doi.org/10.1109/TNSRE.2024.3520720

17. Eng, J. J., and A. M. Pastva. Advances in Remote Monitoring for Stroke Recovery. *Stroke* 53:, 2022. https://doi.org/10.1161/STROKEAHA.122.038885

18. Appleby, E., S. T. Gill, L. K. Hayes, T. L. Walker, M. Walsh, and S. Kumar. Effectiveness of telerehabilitation in the management of adults with stroke: A systematic review. *PLoS One* 14:e0225150, 2019. https://doi.org/10.1371/JOURNAL.PONE.0225150

19. Webster, P. Virtual health care in the era of COVID-19. *The Lancet* 395:1180–1181, 2020. https://doi.org/10.1016/S0140-6736(20)30818-7

20. Wosik, J., M. Fudim, B. Cameron, Z. F. Gellad, A. Cho, D. Phinney, S. Curtis, M. Roman, E. G. Poon, J. Ferranti, J. N. Katz, and J. Tcheng. Telehealth transformation: COVID-19 and the rise of virtual care. *Journal of the American Medical Informatics Association* 27:957–962, 2020. https://doi.org/10.1093/JAMIA/OCAA067

21. Shenoy, A., M. S. Samra, K. Van Ooteghem, K. B. Beyer, S. Thomson, W. E. McIlroy, J. J. Eng, and C. L. Pollock. Evaluating the Usability of Inertial Measurement Units for Measuring and Monitoring Activity Post-Stroke: A Scoping Review. *Sensors 2025, Vol. 25, Page 3694* 25:3694, 2025. https://doi.org/10.3390/S25123694

22. Haeuber, E., M. Shaughnessy, L. W. Forrester, K. L. Coleman, and R. F. Macko. Accelerometer monitoring of home- and community-based ambulatory activity after stroke. *Arch. Phys. Med. Rehabil.* 85:1997–2001, 2004. https://doi.org/10.1016/J.APMR.2003.11.035

23. Rand, D., J. J. Eng, P. F. Tang, J. S. Jeng, and C. Hung. How Active Are People With Stroke? *Stroke* 40:163–168, 2009. https://doi.org/10.1161/STROKEAHA.108.523621

24. Rand, D., and J. J. Eng. Disparity between functional recovery and daily use of the upper and lower extremities during subacute stroke rehabilitation. *Neurorehabil. Neural Repair* 26:76–84, 2012. https://doi.org/10.1177/1545968311408918

25. Halloran, S., L. Tang, Y. Guan, J. Q. Shi, and J. Eyre. Remote monitoring of stroke patients' rehabilitation using wearable accelerometers. *Proceedings - International Symposium on Wearable Computers, ISWC* 72–77, 2019. https://doi.org/10.1145/3341163.3347731

26. Leirós-Rodríguez, R., J. L. García-Soidán, and V. Romo-Pérez. Analyzing the Use of Accelerometers as a Method of Early Diagnosis of Alterations in Balance in Elderly People: A Systematic Review. *Sensors 2019, Vol. 19, Page 3883* 19:3883, 2019. https://doi.org/10.3390/S19183883

27. Shahzad, A., S. Ko, S. Lee, J. A. Lee, and K. Kim. Quantitative Assessment of Balance Impairment for Fall-Risk Estimation Using Wearable Triaxial Accelerometer. *IEEE Sens. J.* 17:6743–6751, 2017. https://doi.org/10.1109/JSEN.2017.2749446

28. Culhane, K. M., M. O'Connor, D. Lyons, and G. M. Lyons. Accelerometers in rehabilitation medicine for older adults. *Age Ageing* 34:556–560, 2005. https://doi.org/10.1093/AGEING/AFI192

29. Vienne, A., R. P. Barrois, S. Buffat, D. Ricard, and P. P. Vidal. Inertial sensors to assess gait quality in patients with neurological disorders: A systematic review of technical and analytical challenges. *Front. Psychol.* 8:817, 2017. https://doi.org/10.3389/FPSYG.2017.00817/FULL

30. Oyake, K., T. Yamaguchi, M. Sugasawa, C. Oda, S. Tanabe, K. Kondo, Y. Otaka, and K. Momose. Validity of gait asymmetry estimation by using an accelerometer in individuals with hemiparetic stroke. *J. Phys. Ther. Sci.* 29:307, 2017. https://doi.org/10.1589/JPTS.29.307

31. Jarchi, D., J. Pope, T. K. M. Lee, L. Tamjidi, A. Mirzaei, and S. Sanei. A Review on Accelerometry-Based Gait Analysis and Emerging Clinical Applications. *IEEE Rev. Biomed. Eng.* 11:177–194, 2018. https://doi.org/10.1109/RBME.2018.2807182

32. Stango, A., F. Negro, and D. Farina. Spatial Correlation of High Density EMG Signals Provides Features Robust to Electrode Number and Shift in Pattern Recognition for Myocontrol. *IEEE Transactions on Neural Systems and Rehabilitation Engineering* 23:189–198, 2015. https://doi.org/10.1109/TNSRE.2014.2366752

33. Ameri, A., M. A. Akhaee, E. Scheme, and K. Englehart. A Deep Transfer Learning Approach to Reducing the Effect of Electrode Shift in EMG Pattern Recognition-Based Control. *IEEE Transactions on Neural Systems and Rehabilitation Engineering* 28:370–379, 2020. https://doi.org/10.1109/TNSRE.2019.2962189

34. Pan, L., D. Zhang, N. Jiang, X. Sheng, and X. Zhu. Improving robustness against electrode shift of high density EMG for myoelectric control through common spatial patterns. *Journal of NeuroEngineering and Rehabilitation 2015 12:1* 12:110-, 2015. https://doi.org/10.1186/S12984-015-0102-9

35. Wang, L., X. Li, Z. Chen, Z. Sun, and J. Xue. Electrode Shift Fast Adaptive Correction for Improving Myoelectric Control Interface Performance. *IEEE Sens. J.* 23:25036–25047, 2023. https://doi.org/10.1109/JSEN.2023.3312403

36. Hu, R., X. Chen, X. Zhang, and X. Chen. Adaptive Electrode Calibration Method Based on Muscle Core Activation Regions and Its Application in Myoelectric Pattern Recognition. *IEEE Transactions on Neural Systems and Rehabilitation Engineering* 29:11–20, 2021. https://doi.org/10.1109/TNSRE.2020.3029099

37. Li, X., A. Holobar, M. Gazzoni, R. Merletti, W. Z. Rymer, and P. Zhou. Examination of poststroke alteration in motor unit firing behavior using high-density surface EMG decomposition. *IEEE Trans. Biomed. Eng.* 62:1242–1252, 2015. https://doi.org/10.1109/TBME.2014.2368514

38. Martinez-Valdes, E., F. Negro, C. M. Laine, D. Falla, F. Mayer, and D. Farina. Tracking motor units longitudinally across experimental sessions with high-density surface electromyography. *J. Physiol.* 595:1479, 2017. https://doi.org/10.1113/JP273662

39. Yokoyama, H., A. Sasaki, N. Kaneko, A. Saito, and K. Nakazawa. Robust Identification of Motor Unit Discharges from High-Density Surface EMG in Dynamic Muscle Contractions of the Tibialis Anterior. *IEEE Access* 9:123901–123911, 2021. https://doi.org/10.1109/ACCESS.2021.3107283

40. Wang, W., K. Li, S. Yue, C. Yin, and N. Wei. Associations between lower-limb muscle activation and knee flexion in post-stroke individuals: A study on the stance-to-swing phases of gait. *PLoS One* 12:e0183865, 2017. https://doi.org/10.1371/JOURNAL.PONE.0183865

41. Gao, F., and L. Q. Zhang. Altered contractile properties of the gastrocnemius muscle poststroke. *https://doi.org/10.1152/japplphysiol.90930.2008* 105:1802–1808, 2008. https://doi.org/10.1152/JAPPLPHYSIOL.90930.2008

42. Silva, A., A. S. P. Sousa, R. Pinheiro, J. Ferraz, J. M. R. S. Tavares, R. Santos, and F. Sousa. Activation timing of soleus and tibialis anterior muscles during sit-to-stand and stand-to-sit in post-stroke vs. healthy subjects. *Somatosens. Mot. Res.* 30:48–55, 2013. https://doi.org/10.3109/08990220.2012.754755

43. Maguire, C., J. M. Sieben, M. Frank, and J. Romkes. Hip abductor control in walking following stroke -- the immediate effect of canes, taping and TheraTogs on gait. *Clin. Rehabil.* 24:37–45, 2010. https://doi.org/10.1177/0269215509342335

44. Chan, P. P., J. I. Si Tou, M. M. Tse, and S. S. Ng. Reliability and Validity of the Timed Up and Go Test With a Motor Task in People With Chronic Stroke. *Arch. Phys. Med. Rehabil.* 98:2213–2220, 2017. https://doi.org/10.1016/J.APMR.2017.03.008

45. Johansen, K. L., R. D. Stistrup, C. S. Schjøtt, J. Madsen, and A. Vinther. Absolute and Relative Reliability of the Timed 'Up & Go' Test and '30second Chair-Stand' Test in Hospitalised Patients with Stroke. *PLoS One* 11:e0165663, 2016. https://doi.org/10.1371/JOURNAL.PONE.0165663

46. GlobalSearch - Find global minimum - MATLABat <https://www.mathworks.com/help/gads/globalsearch.html>

47. Ugray, Z., L. Lasdon, J. Plummer, F. Glover, J. Kelly, and R. Marti. Scatter Search and Local Nlp Solvers: A Multistart Framework for Global Optimization. *SSRN Electronic Journal* , 2006. https://doi.org/10.2139/SSRN.886559

48. Hussain, I., and S. J. Park. Prediction of Myoelectric Biomarkers in Post-Stroke Gait. *Sensors 2021, Vol. 21, Page 5334* 21:5334, 2021. https://doi.org/10.3390/S21165334

49. Toral, V., F. J. Romero, E. Castillo, D. P. Morales, A. Rivadeneyra, A. Salinas-Castillo, L. Parrilla, and A. García. A versatile wearable based on reconfigurable hardware for biomedical measurements. *Measurement* 201:111744, 2022. https://doi.org/10.1016/J.MEASUREMENT.2022.111744

50. Chen, C., G. Chai, W. Guo, al -, Y.-X. Zhou, H.-P. Wang, X.-L. Bao, G. Marco, B. Alberto, and V. Taian. Surface EMG and muscle fatigue: multi-channel approaches to the study of myoelectric manifestations of muscle fatigue. *Physiol. Meas.* 38:R27, 2017. https://doi.org/10.1088/1361-6579/AA60B9

51. Cumming, T. B., M. Packer, S. F. Kramer, and C. English. The prevalence of fatigue after stroke: A systematic review and meta-analysis. *https://doi.org/10.1177/1747493016669861* 11:968–977, 2016. https://doi.org/10.1177/1747493016669861

52. De Groot, M. H., S. J. Phillips, and G. A. Eskes. Fatigue associated with stroke and other neurologic conditions: implications for stroke rehabilitation. *Arch. Phys. Med. Rehabil.* 84:1714–1720, 2003. https://doi.org/10.1053/S0003-9993(03)00346-0

53. Cifrek, M., V. Medved, S. Tonković, and S. Ostojić. Surface EMG based muscle fatigue evaluation in biomechanics. *Clinical Biomechanics* 24:327–340, 2009. https://doi.org/10.1016/J.CLINBIOMECH.2009.01.010

54. Negro, F., K. E. Bathon, J. N. Nguyen, C. G. Bannon, C. Orizio, S. K. Hunter, and A. S. Hyngstrom. Impaired Firing Behavior of Individually Tracked Paretic Motor Units During Fatiguing Contractions of the Dorsiflexors and Functional Implications Post Stroke. *Front. Neurol.* 11:540893, 2020. https://doi.org/10.3389/FNEUR.2020.540893/BIBTEX

55. Bigland, B., and O. C. J. Lippold. The relation between force, velocity and integrated electrical activity in human muscles. *J. Physiol.* 123:214–224, 1954. https://doi.org/10.1113/JPHYSIOL.1954.SP005044

56. Hof, A. L. EMG and muscle force: An introduction. *Hum. Mov. Sci.* 3:119–153, 1984. https://doi.org/10.1016/0167-9457(84)90008-3

57. Disselhorst-Klug, C., T. Schmitz-Rode, and G. Rau. Surface electromyography and muscle force: Limits in sEMG–force relationship and new approaches for applications. *Clinical Biomechanics* 24:225–235, 2009. https://doi.org/10.1016/J.CLINBIOMECH.2008.08.003

58. Lai, A. K. M., G. A. Lichtwark, A. G. Schache, and M. G. Pandy. Differences in in vivo muscle fascicle and tendinous tissue behavior between the ankle plantarflexors during running. *Scand. J. Med. Sci. Sports* 28:1828–1836, 2018. https://doi.org/10.1111/sms.13089

59. Kluding, P. M., K. Dunning, M. W. O'Dell, S. S. Wu, J. Ginosian, J. Feld, and K. McBride. Foot drop stimulation versus ankle foot orthosis after stroke: 30-week outcomes. *Stroke* 44:1660–1669, 2013. https://doi.org/10.1161/STROKEAHA.111.000334;PAGE:STRING:ARTICLE/CHAPTER